\documentclass[manuscript]{acmart}
\usepackage{natbib}

\AtBeginDocument{%
  \providecommand\BibTeX{{%
    \normalfont B\kern-0.5em{\scshape i\kern-0.25em b}\kern-0.8em\TeX}}}

\setcopyright{acmlicensed}

\copyrightyear{2024}
\acmYear{2024}
\setcopyright{acmlicensed}\acmConference[CHI '24]{Proceedings of the CHI Conference on Human Factors in Computing Systems}{May 11--16, 2024}{Honolulu, HI, USA}
\acmBooktitle{Proceedings of the CHI Conference on Human Factors in Computing Systems (CHI '24), May 11--16, 2024, Honolulu, HI, USA}
\acmDOI{10.1145/3613904.3642392}
\acmISBN{979-8-4007-0330-0/24/05}

\begin{document}

\title{A Human-Centered Review of Algorithms in Homelessness Research}


\author{Erina Seh-Young Moon}
\affiliation{%
  \institution{University of Toronto}
  \city{Toronto}
  \country{Canada}}
\email{erina.moon@mail.utoronto.ca}

\author{Shion Guha}
\affiliation{%
  \institution{University of Toronto}
  \city{Toronto}
  \country{Canada}}
\email{shion.guha@utoronto.ca}

\renewcommand{\shortauthors}{Moon and Guha}

\begin{abstract}
Homelessness is a humanitarian challenge affecting an estimated 1.6 billion people worldwide. In the face of rising homeless populations in developed nations and a strain on social services, government agencies are increasingly adopting data-driven models to determine one’s risk of experiencing homelessness and assigning scarce resources to those in need. We conducted a systematic literature review of \textbf{57} papers to understand the evolution of these decision-making algorithms. We investigated trends in computational methods, predictor variables, and target outcomes used to develop the models using a human-centered lens and found that only \textbf{9} papers (15.7\%) investigated model fairness and bias. We uncovered tensions between explainability and ecological validity wherein predictive risk models (53.4\%) unduly focused on reductive explainability while resource allocation models (25.9\%) were dependent on unrealistic assumptions and simulated data that are not useful in practice. Further, we discuss research challenges and opportunities for developing human-centered algorithms in this area.  
\end{abstract}




\maketitle

\section{Introduction}

Homelessness poses a persistent challenge in many nations worldwide with an estimated 1.6 billion people living in inadequate housing \cite{un}. In recent years, homelessness has become a rising challenge in the United States and Canada \cite{wsj_trend, canada} with major cities declaring states of emergency in response to this humanitarian crisis \cite{la, ny, Toronto}. Homelessness is deeply intertwined and often precipitated by poverty, affordable housing shortages, medical issues, and involvement with criminal justice or child welfare services \cite{ chelmis_21, byrne_classification_2020, shinn1998}. Unhoused individuals face significant challenges as they seek necessities (i.e., food and shelter) and are often stigmatized and discriminated against when accessing health care services and employment opportunities \cite{ecology2010}.

Over the last three decades, countries have increasingly adopted a data-driven approach in response to this humanitarian plight. The Housing First policy was adopted in many jurisdictions following empirical findings that unconditionally housing individuals can alleviate chronic homelessness and reduce public costs \cite{Stanhope_Dunn_2011}. Moreover, the US and Canada have federally mandated data-driven centralized systems to assess needs and efficiently allocate services to those with the greatest need \cite{Ecker2022}. In continuation of this trend, the US and Canada have begun expanding the use of automated decision-making algorithms in the public sector with the homelessness domain being no exception \cite{levy21, canada_algs}. Within homelessness, government agencies have increasingly adopted algorithms to predict one’s risk of experiencing homelessness or related harms \cite{Wachter_19, toronto_algorithm, toros18, kithulgoda_predictive_2022}.

HCI scholars have been at the forefront of examining algorithmic systems in high-stakes decision-making environments such as homelessness, education, child welfare, and online behavior with an interest in understanding the social impact of these algorithm \cite{saxena2020human, kelly23, razi21, kim21_cyber, showkat23, karusala19, kuo23, clancy2022}. Works have examined potential harms and limitations of these systems and investigated issues around bias and fairness by incorporating stakeholder perceptions \cite{feng22, saxena2023algorithmic, lee2017algorithmic}. Specific to algorithms in homelessness research, recent SICGHI work has uncovered the need to adopt human-centered approaches to the algorithm design process. Karusala et al. \cite{karusala19} and Kuo et al. \cite{kuo23} studied stakeholder perceptions of algorithms used in homelessness and found it failed to account for contextual case details and interactive dynamics between caseworkers and clients. Showkat et al. \cite{showkat23} studied values encoded in homelessness algorithms, and found human context became lost through computational abstractions (i.e., data preprocessing and model parameter tuning). As public sector agencies are increasingly adopting data-driven systems in homelessness and SIGCHI continues to pursue research on the human impact of decision-making algorithms in homelessness, we need to critically examine the technical underpinnings of the algorithms to understand how algorithm design choices impact those who regularly interact with the models and identify research gaps. Synthesizing existing decision-making algorithms in homelessness research can help address this need. Thus, in our study, we adopt a human-centered algorithm design lens \cite{baumer2017toward} and interrogate the impact of models’ design configurations by asking the following high-level research questions:

\vspace{0.1cm}
\begin {itemize} 
  \item \textbf{RQ1:} \textit{What computational methods are employed when building algorithms for homelessness?}
  \item \textbf{RQ2:} \textit{What predictor variables are used to build algorithms for homelessness?}
  \item \textbf{RQ3:} \textit{What target outcomes do algorithms for homelessness predict?}
\end{itemize}

To answer these questions, we conducted a systematic literature review of algorithms in homelessness research from 55 peer-reviewed papers and 2 white papers published from 1998 to 2023. We qualitatively analyzed the current design of these algorithms using a human-centered algorithm design lens \cite{baumer2017toward} to assess the functionality and ecological validity of the models \cite{Ecker2022, fowlersolving2019}. Our results revealed that only nine papers (15.7\% of corpus) adopted a human-centered perspective to interrogate issues of fairness, justice, and bias in algorithms that challenge current approaches to support the unhoused. Moreover, our findings uncovered tensions between algorithmic explainability and real world context. Predictive risk models (53.4\% of corpus) could generate explainable outcomes by focusing on predicting ‘risk’ using a narrow set of easily quantifiable administrative data. These models primarily prioritized \textit{who} should receive services without fully accounting for \textit{what} a person needs and the ecological landscape of homelessness. At the same time, we identified an increasing interest in resource allocation algorithms in the last five years (25.9\%). While these models recognized the practical challenges of allocating resources to unhoused persons within a resource-constrained setting, they were limited in applicability due to them being mostly opaque simulation systems. In summary, we make the following research contributions:
\begin {itemize} 
\item We critically review trends in the proposed designs for algorithms for homelessness using a human-centered algorithm design theoretical lens \cite{baumer2017toward}.

\item We question the feasibility and appropriateness of designing algorithms for homelessness \cite{baumer11}. 
\item We identify gaps and future research opportunities for unhoused human-centered algorithm design.    

\end{itemize}

In the next section, we discuss the human-centered algorithm conceptual framework and review existing literature on algorithmic design in homelessness. Then, we explain our methods followed by our findings and discussion.

\section{Related Work}

The SIGCHI community has been at the forefront of conducting public sector research to design sociotechnical systems that uplift vulnerable populations, including children, the unhoused, migrant youths, and asylum seekers \cite{saxena2021framework2, ledantec2009, oghenemaro23, tachtler21}. SIGCHI also has a tradition of engaging with the unhoused to promote empowerment using participatory methods. Woelfer and Hendry \cite{Woelfer_Hendry_2012} conducted interviews with the unhoused to understand how their usage of social network sites inform their identities and social ties. Halperin et al. \cite{halperin23} also recently introduced a design of a conversational storytelling agent that supports the documentation of the lived experiences of the unhoused. Relevant to our work, recent SIGCHI research on algorithmic systems in homelessness has adopted a human-centered lens to deeply interrogate the social and human impact of- and values embedded in decision-making homelessness algorithms. Kuo et al. \cite{kuo23} introduced a comic boarding method to elicit stakeholder perspectives for an algorithmic system in homelessness services. In the study, frontline caseworkers and unhoused individuals expressed concern about the utility of using administrative data to train and build the model and questioned the construct validity of proxy outcomes used by the algorithm. In a related vein, Showkat et al. \cite{showkat23} examined values embedded in homelessness algorithms and found contextual circumstances of unhoused persons were lost through the machine learning model development pipeline. Through investigations of stakeholder perspectives and values in these algorithms, SIGCHI scholars have highlighted the need to center human perspectives in the design of homelessness algorithms.

Outside of SIGCHI, there have been comparatively few works focused on designing and considering the social impact of implementing street-level decision making algorithms for the unhoused. Instead, there has been a wealth of computational work studying common characteristics of the unhoused and identifying risk factors of homelessness. For example, Roebuck et al. \cite{roebuck2023}, Shelton et al. \cite{Shelton09}, and Wong et al. \cite{wong97} used regression models to determine risk factors associated with homelessness. Culhane and Kuhn \cite{kuhn98} used cluster analysis to identify three types of shelter users (transitional, episodic, and chronic). Many of these works have investigated homeless risk factors associated with specific unhoused groups such as families \cite{shinn2013}, youth \cite{tabar20}, and veterans \cite{Byrne2016_veteran} to address their needs.

Our study builds on human-centered SIGCHI work by Kuo et al. \cite{kuo23} and Showkat et al. \cite{showkat23}, which calls to the forefront, the need to empower human perspectives in homelessness algorithm design. One way to address this need is to critically examine the design of these models through a review of literature on decision-making algorithms in homelessness research using a human-centered lens. Baumer’s human-centered algorithm design (HCAD) framework \cite{baumer2017toward} offers a theoretical lens to critically examine the design of such algorithms by examining the (often-ignored) social interpretations of algorithms by lay persons. Baumer \cite{baumer2017toward} recommends implementing HCAD through three strategies: (1) the \textit{theoretical strategy} to incorporate domain-specific theory into algorithm predictor selection, evaluation, and dataset selection to build the models; (2) the \textit{participatory strategy} that involves stakeholders in the design process to bridge the gap between algorithm developers and impacted stakeholders; and (3) the \textit{speculative strategy} that provokes dialogue around benefits, unintended consequences, and future algorithm design opportunities.

Recent SIGCHI work has shown Baumer’s HCAD lens can be applied in literature review studies to unveil trends, gaps, and opportunities for further algorithmic research in child welfare, higher education, online sexual risk detection, and cyberbullying detection \cite{saxena2020human, kelly23, razi21, kim21_cyber}. Close to our work, Saxena et al. \cite{saxena2020human} and McConvey et al. \cite{kelly23} conducted literature reviews using an HCAD lens for decision-making algorithms in the public sector. The authors asked what predictors, computational methods, and target outcomes were used in child welfare and higher-education decision-making algorithms, respectively. Through this framework, Saxena et al. identified a need to shift towards strength-based models because the algorithms heavily focused on assessing a child’s maltreatment risk using case risk factors as predictors. McConvey et al. found that higher-education algorithms were increasingly built using less explainable methods and student-protected attributes as predictors (e.g., race, disability status), which could cause algorithmic harms. Given the versatility of Baumer’s HCAD lens \cite{baumer2017toward} to critically examine the human and social impact of decision-making algorithms, we ask: what computational methods, predictor variables, and target outcomes are employed in algorithms for homelessness to unveil critical insights in this research space.

\section{Research Context}

In this section, we provide background information about data-driven practices in homelessness. As most papers in our corpus introduced models within the US and Canadian context (n=55, 96.5\%), we provide background information for these countries. In recognition of the resource-constrained landscape of homelessness and to efficiently allocate services to individuals based on their needs, the US and Canada have federally mandated the implementation of coordinated systems to end homelessness \cite{openingdoors, reachinghomes, Ecker2022}. Recognizing that different communities face unique challenges, the US Department of Housing and Urban Development and the Canadian government through the Reaching Home Directives have set forward minimum requirements on implementing coordinated systems while allowing for flexibility in how communities choose to tailor coordinated systems to their local context \cite{Ecker2022, coordinatedguide, reachinghomesdirectives, hudselfassess}.

At a minimum, coordinated systems operate through four pillars: access, assessment, prioritization, and matching/referral \cite{Ecker2022}. All individuals must be able to equitably access entry into homeless support systems through clear access points. Individuals seeking services must be assessed on their level of need using standardized assessment tools. Assessment scores, along with other information on clients, are then used to determine a person’s priority for homeless support services, and the client is then matched and referred to a service through the centralized system \cite{openingdoors, reachinghomes}. Communities must also build a governance operating model that defines policies and procedures on how the coordinated systems operate \cite{coordinatedguide, reachinghomesdirectives, hudselfassess}. This includes determining which assessment tools should be used to assess client needs, a list of prioritization factors, service referral procedures, and protocols to obtain, retain, and share client information. Communities thus have considerable latitude on what and how computational decision-making tools are used within homeless services. To cite an example of the diverse approaches to tackle homelessness in Canada; London, Canada \cite{VanBerlo2021, london}, uses an AI prediction tool to predict one’s likelihood of experiencing chronic homelessness, while Toronto, Canada, does not and uses a triage tool for needs assessment \cite{torontostar}.

Before implementing coordinated systems, homeless support services were often offered on a first-come, first serve basis \cite{Ecker2022}, and service providers competed for limited funding and available rooms \cite{ eubanks2018automating}. Although coordinated systems have become a central component of the homelessness system and is the current approach to homelessness, there is debate about whether it leads to equitable and consistent outcomes for the unhoused \cite{ eubanks2018automating, karusala19}. Nevertheless, current algorithms in homelessness must operate within these systems, and we must consider this sociotechnical context when designing these models. Therefore, using a human-centered centered-lens, we conducted a systematic literature review of algorithms in homelessness and assessed their ecological validity.

\section{Methods}

\subsection{Scoping criteria and data collection process}

In this study, we define algorithms using Alkhatib and Bernstein's \cite{alkhatib2019street} conceptualization of \textit{street-level algorithms}. Street-level algorithms are computational systems that use programmed rules to interact and make on-the-ground street-level decisions about people within sociotechnical systems \cite{alkhatib2019street}. With this definition of algorithms in mind, we followed guidelines proposed by Webster and Watson \cite{webster2002analyzing} to direct our systematic literature review. The unit of analysis for this literature review was the algorithm described in the article. To identify papers describing prior, current, and new algorithms being proposed by researchers, we used the following search terms: "homeless," "homelessness," "unhoused," "optimization," "resource allocation," "AI," "neural network," "machine learning," "regression," "housing insecurity," "computation," "data-driven," and "algorithm." We show the combination of search terms we used in Table \ref{tab:searchterms} in the Appendix. We set the following inclusion criteria (any articles that did not meet these items were excluded from our review).

\begin {itemize} 
  \item Papers are peer-reviewed published work or reports published by governmental agencies written in English.
  \item The study/report discusses technical components of the algorithm, including computational methods used and model predictors and target outcomes.
\end{itemize}

We searched in the ACM Digital Library, IEEE Xplore, Springer, Routledge, and Elsevier to collect articles from multiple disciplines. We placed no time period constraints in the selection process. We also cross-referenced citations in each paper to identify and include literature that met our inclusion criteria. We identified 57 relevant articles that met our inclusion criteria. Articles were situated in the US (n=49), Canada (n=6), Spain (n=1), and Australia (n=1). We include the flow diagram showing our literature review process in the Appendix. 

\subsection{Data analysis}
We conducted a structured qualitative analysis using a grounded thematic process \cite{braun2006using}. The first author closely read each paper with a particular focus on the abstract, methods, and results to generate codes as seen in Table \ref{tab:codebook_summary}. We also coded for model descriptive characteristics seen in Table \ref{tab:descriptive}. Many papers included multiple computational methods, so we coded multiple methods for each model. When coding the different computational methods, we differentiated machine learning methods (e.g., random forest, decision trees) from statistical models (e.g., generalized linear models) based on Breiman's assumptions around data \cite{breiman2001statistical}. Only 1 paper described 2 different algorithms with a different set of predictors and target outcomes, and other papers described 1 algorithm each. The first author consulted with the coauthor to reach a consensus around codes at the beginning and during the coding process to resolve ambiguous codes.

\section{Results}

In this section, we present the results of our literature review. We first outline the descriptive characteristics of our dataset, followed by a description of our results for our research questions. Lastly, we investigate the relationships between the predictors, outcome variables, and computational methods.

\subsection{Dataset Characterstics}
We identified fifty-seven papers that met our scoping criteria. All papers described one algorithm each except for one paper \cite{toros19} which described two algorithms. Most papers in our dataset were published in computer science (n=36) and social science venues (n=12). Other papers were published in health venues (n=8) and in an engineering venue \cite{kayaimproving_2022}. All academic papers included in our dataset (n=55) were peer-reviewed and were published in journals (n=29) and conferences (n=26). The white papers in our dataset (n=2) were published by research organizations: the Office of Policy Development and Research (PD\&R) of the U.S. Department of Housing and Urban Development \cite{toros18} and the Economic Roundtable \cite{toros19}. Some papers introduced algorithms designed specifically for unhoused youth (n=19), families (n=8), veterans (n=2), patients (n=2), those experiencing substance use disorders (n=1), and only adults (n=1) as researchers argued different unhoused subpopulation groups had different needs \cite{kuhn98}.

\begin{table}[]
\centering
\begin{tabular}{l|c|l}
\hline
\textbf{Classification} & \textbf{n} & \textbf{Breakdown} \\ \hline
Peer Reviewed & 55 & CS (36); social science (10); health (8); engineering (1) \\ \hline
Agency Report & 2 & Social Science (2) \\ \hline
Journal & 29 & CS (11); social science (9); health (8); engineering (1) \\ \hline
Conference & 26 & CS (25); social science (1) \\ \hline
Study Focus & 58 & General population (25); youth (19); families (8); veterans (2); \\
 &  & patients (2); substance use disorder (1); adults (1) \\ \hline
\end{tabular}
\caption{Dataset Descriptive Characteristics (56 papers describe one algorithm each but one paper \cite{toros19} details two algorithms which is why study focus n=58 and not 57.)}
\label{tab:descriptive}
\end{table}

\subsection{Computational Methods Employed in Algorithms for Homelessness (RQ1)} \label{sec:rq1}

This section discusses the computational methods employed to develop algorithms for homelessness organized by the categories: inferential statistics, machine learning, deep learning, and optimization, as seen in Table \ref{tab:codebook_summary}. 

\subsubsection{\textit{Inferential Statistics}} \label{sec:infstat}

Figure \ref{fig:method_time} shows inferential statistics was the most consistently popular method over time in our dataset: 35 models (60.3\%) employed generalized linear models (GLM) and six models employed statistical tests (10.3\%), including chi-squared and t-tests, to select and understand predictors of interest when running GLMs \cite{balagot2019, moxley2020, toros18, alcalde2022}. The logistic regression model was the most common GLM used (n=25, 43.1\%). Cox regression (n=3) and Bayesian additive regression trees (BART) (n=6) were also used. Our findings highlight limitations in the current design of homelessness algorithms. Historically, computational homelessness research often used GLMs to identify common characteristics (or types) of unhoused individuals \cite{wong97, roebuck2023, padgett1995, jadidzadeh2019, hsu2021, hanauer2021, diGuiseppi2020}. Our findings showed that this popular methodology has been extended to designing street-level homelessness algorithms due to their explainability. Some papers in our corpora explained they used GLMs because of their ability to (1) predict outcomes, (2) explain how predictors influence the probability of the outcome variable, and (3) garner stakeholder trust due to their explainability \cite{vanberlo21, hong2018, kube_fair_2023}. While GLMs carry these advantages, recent work by Saxena et al. \cite{saxena2020human}, who studied the use of GLMs to build algorithms in an adjacent domain- child welfare- has warned that GLMs can raise ethical issues when used for individual-level decision-making as they inherently predict/observe average, macro-level patterns while excluding outlier cases \cite{stevens1984}. These criticisms also apply in algorithms for homelessness. An algorithm's ability to provide a transparent explanation of how a GLM rated an individual’s risk of experiencing chronic homelessness could be at the cost of the algorithm generating an ill-informed outcome by ignoring temporal, contextual factors that do not fit the ‘average’ trend. For example, an individual with a history of substance abuse may be deemed high 'risk' by a GLM but they may have a strong will to be independent and be housed that will help them overcome challenges \cite{karusala19}.

\begin{table}[]
\centering
\resizebox{\textwidth}{!}{%
\begin{tabular}{l|l|l|c|c|c}
\textbf{} & \textbf{Dimension} & \textbf{Code} & \textbf{Count} & \textbf{\%} & \textbf{Example} \\ \hline
\textbf{Computational} & \textbf{Inferential Statistics} & Statistical tests (ST) & 6 & 10.3\% &  \cite{toros18}\\
\textbf{Method} & \textbf{} & Generalized Linear Models (GLM) & 35 & 60.3\% &  \cite{moore_prospective_2012}\\
\textbf{} & \textbf{Machine Learning} & Supervised Learning (SUP) & 19 & 32.8\% &  \cite{chelmis21}\\
\textbf{} & \textbf{} & Unsupervised  Learning (USUP) & 2 & 3.4\% &  \cite{hong2018} \\
\textbf{} & \textbf{} & Natural Language Processing (NLP) & 1 & 1.7\% &  \cite{dou_harnessing_2021}\\
\textbf{} & \textbf{Deep Learning} & Neural Network (NN)& 5 & 8.6\% &  \cite{messier_predicting_2022}\\
\textbf{} & \textbf{Optimization} & Mixed Integer Programming (MIP) & 7 & 12.1\% &  \cite{rahmattalabi2022}\\
\textbf{} & \textbf{} & Network Analysis (NA) & 8 & 13.8\% &  \cite{young2020_dynamism}\\ 
\textbf{} & \textbf{} & System Simulation (SS) & 7 & 12.1\% &  \cite{fowler_meeting_2022}\\ 
\textbf{} & \textbf{} & Heuristic Algorithm (HA) & 2 & 3.4\% &  \cite{khayyatkhoshnevis_smart_2020}\\ \hline
\textbf{Predictor} & \textbf{Demographics} & Individual demographics & 37 & 63.8\% &  \cite{hamilton_risk_2022}\\
\textbf{Variables} & \textbf{} & Household composition & 11 & 19.0\% &  \cite{shinn2013}\\
\textbf{} & \textbf{} & Economic & 24 & 41.4\% &  \cite{purao2019}\\
\textbf{} & \textbf{Housing} & Current Housing & 22 & 37.9\% &  \cite{toros19}\\
\textbf{} & \textbf{} & Housing History & 22 & 37.9\% &  \cite{chan_evidence_2017}\\
\textbf{} & \textbf{} & Service usage history & 25 & 43.1\% &  \cite{greer2016}\\
\textbf{} & \textbf{} & Housing/service needs & 11 & 19.0\% &  \cite{kaya_discrete_2022} \\
\textbf{} & \textbf{Services} & Service provider information & 20 & 34.5\% &  \cite{kayaimproving_2022}\\
\textbf{} & \textbf{Health} & Health  & 36 & 62.1\% &  \cite{byrne_predictive_2019} \\
\textbf{} & \textbf{Person Needs/Risks} & Prior victimization/trauma & 20 & 34.5\% &  \cite{azizi2018}\\
\textbf{} & \textbf{} & Involvement with criminal justice & 16 & 27.6\% &  \cite{farrell_reassessing_2023}\\
\textbf{} & \textbf{} & Risk assessment & 11 & 19.0\% &  \cite{mullen2022}\\
\textbf{} & \textbf{} & Behavioral Characteristics & 2 & 3.4\% &  \cite{yadav_optimal21}\\
\textbf{} & \textbf{Relationships} & Relationship Strengths & 7 & 12.1\% &  \cite{shinn1998}\\
\textbf{} & \textbf{} & Social Network Analysis & 8 & 13.8\% &  \cite{yadav2016}\\ \hline
\textbf{Outcome} & \textbf{Outcome} & Resource Allocation (RES) & 15 & 25.9\% &  \cite{kube_just_2022} \\
\textbf{Variables} & \textbf{} & Risk of experiencing homeless-related harms (HARM) & 8 & 13.8\% &  \cite{tabar20}\\
\textbf{} & \textbf{} & Influential person identification (ID) & 7 & 12.1\% &  \cite{srivastava2019}\\
\textbf{} & \textbf{} & Predict next trajectory (TRJ) & 6 & 10.3\% &  \cite{rahman2022} \\
\textbf{} & \textbf{} & Risk of homelessness (RISK) & 23 & 39.7\% & \cite{koh2022}
\end{tabular}%
}
\caption{Codebook organized by research question(Percentages shown out of n=58 algorithms in our corpus of 57 papers)}
\label{tab:codebook_summary}
\end{table}

\subsubsection{Machine Learning (ML) and Deep Learning (DL) Methods}

\begin{figure}[h]
  \centering
  \includegraphics[scale=0.55]{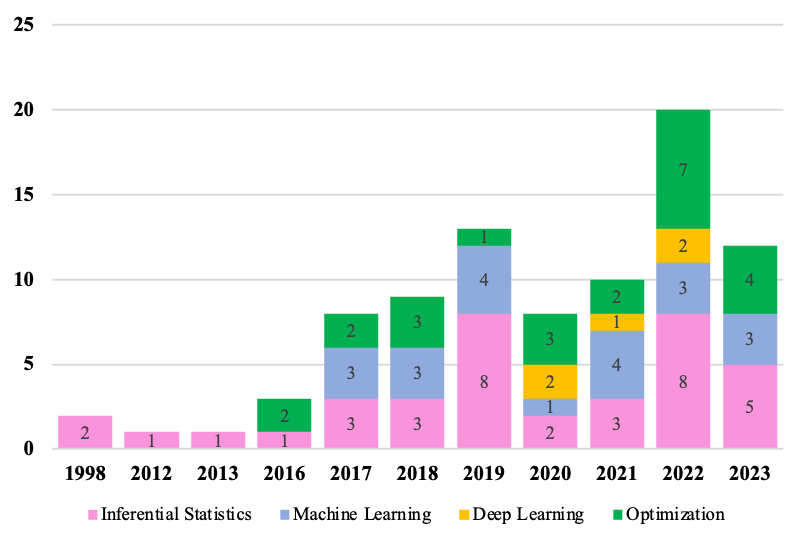}
  \caption{The methods used to build algorithms}
  \label{fig:method_time}  
\end{figure}

We found fewer instances of machine learning (ML) and deep learning (DL) approaches in our dataset: 36.2\% (n=21) of models used ML techniques, and 8.6 \% (n=5) models employed deep learning (DL) methods. Figure \ref{fig:method_time} shows that ML algorithms have been used consistently since 2017— mostly incorporating supervised ML methods (n=19, 32.8\%). Homeless systems often involve regulatory practices where accountability and transparency on how allocation decisions are made are emphasized \cite{chelmis_21}. In accordance with these demands and resonating with the reasons for choosing GLMs, we observed ML approaches that were more ‘explainable’ were favored among ML and DL approaches. Random forest (n=12, 20.7\%) and decision trees (n=8, 13.8\%) were the most popular ML methods, with several studies in our dataset citing their ability to determine feature importance \cite{kube_fair_2023, alcalde2022, gao17} and explain the classification process \cite{toros18} as reasons for choosing the models. DL methods that are often considered opaque systems were only used in 5 algorithms (8.6\%) in our dataset and were used to predict the future state for unhoused persons \cite{fisher22, vanberlo21, fisher_simulating_2020}. While a variety of ML techniques, such as support vector machines, K-nearest neighbors, and gradient-boosted trees were also used to design algorithms in our dataset, we noticed most of these approaches were supervised ML approaches. Supervised ML is well suited for classification tasks. Our findings showed supervised ML was used primarily to profile ‘who’ is at risk of experiencing homelessness (explained further in Section \ref{sec:mp}). Even the two models in our corpora that used unsupervised ML methods (K-means clustering) focused on identifying service provider clusters \cite{tsai17} and different types of at-risk families entering the shelter system \cite{hong2018}. Notwithstanding the ethical issues that arise from typecasting individuals and potentially reinforcing historical injustices \cite{showkat23, eubanks2018automating, kuo23}, research on homelessness has long found that homelessness is caused, sustained and resolved by a confluence of individual, community, and societal-level factors such as availability of public benefits and family conflict/abuse \cite{ecology2010, piat2015}. Thus, algorithms that seek to categorize individuals may merely provide a narrow snapshot of an individual’s circumstances without fully considering the community and welfare systems they live in.

\subsubsection{Optimization Techniques} \label{sec:ot} 
We saw optimization approaches quickly became the second most popular method in our dataset since 2016 (n=24, 41.4\%), rivaling inferential statistics’ popularity in 2022 (see Figure \ref{fig:method_time}). By design, optimization methods offer advantages as they offer an algorithmic methodology to consider the temporally dynamic ecology of homeless services. Notably, in our dataset, network analysis algorithms (n=8, 13.8\%) expressed networks as social connections between unhoused youth or as a series of interconnected services offered within the homeless system to deploy community-driven interventions or allocate resources efficiently. System simulation models (n=7, 12.1\%) and mixed integer programming algorithms (n=7, 12.1\%) in our corpus were used to track the flow of resources between different homeless service providers based on their capacity given unhoused person counts/needs. 

The rise of optimization approaches coincides with the implementation of federally mandated centralized homeless support systems since 2012 (i.e., coordinated system) in the US and Canada \cite{hud21, caeh18}. A centralized homeless support system means collecting real-time data on clients and homeless services to prioritize support to those with the greatest need \cite{hud21, caeh18}. Optimization approaches can account for the dynamic interactions in centralized systems between individuals and homeless \textit{system-level} components such as homeless support services to optimize outcomes (e.g., resource allocation, minimizing violence). This way, they can shift the focus from identifying/categorizing ‘who’ unhoused clients are to considering ‘what' services are available to them within the homeless support system \cite{ecology2010}. While these approaches can theoretically provide greater reliability as they can consider a wider breadth of factors impacting unhoused clients compared to GLMs or supervised ML algorithms identified above, we also found that most optimization models in our dataset were simulation studies that lacked external validity. They used administrative or survey data - data that is often sparse and reductive \cite{chelmis_21, saxena2021framework2} or randomly generated data (i.e., in heuristic algorithms) – to build and run simulations.

\subsection{Predictors Deployed in Algorithms for Homelessness}

In this section, we examine the predictors employed in our corpus. As algorithms employed varying numbers of different predictors, we coded the predictors by the dimensions, \textbf{Demographics, Housing, Services, Health, Person Needs/Risks}, and \textbf{Relationships} (see Table \ref{tab:codebook_summary}). 

\subsubsection{Predictors about demographics and health-related variables}

\begin{figure}[]
\centering 
\includegraphics[scale=0.55]{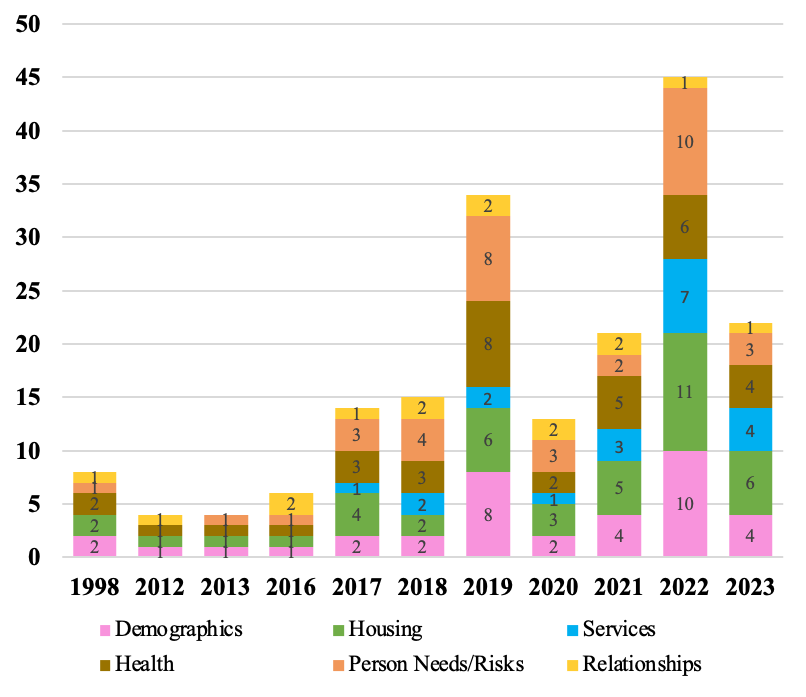}
\caption{The predictors used in the algorithms}
\label{fig:predictor_time}
\end{figure}

We observed demographic and health-related predictors have been a mainstay in algorithms in homelessness (see Figure \ref{fig:predictor_time}). \textbf{Demographics} (n=37, 63.8\%) provided signals on ‘who’ someone is based on their (1) \textit{Individual demographics} (e.g., age, gender, education level) (n=37, 63.8\%), (2) \textit{Household composition} (e.g., number of adults/children) (n=11, 19.0\%), and (3) \textit{Economic}-related predictors (e.g., monthly income) (n=24, 41.4\%). \textbf{Health} (n=36, 62.1\%) predictors provided some information on one's needs related to whether an individual or family member had a disabling, physical, developmental, mental health, and substance-abuse problems and their history of medical treatment. Our findings are important because first, we found most algorithms used easily quantifiable data from public administrative records to build their models. While these are convenient inputs to describe ‘who’ someone is, they cannot account for difficult-to-quantify (but critical) information on individuals, such as information on whether they have a support system among close relatives/friends who can help them overcome challenges. Second, we found the focus on using demographic and health-related predictors goes against existing homelessness research, which has long-refuted the assumption that individual-level and health-related factors are the main determinants of homelessness \cite{ecology2010, Fischer_Breakey_1991, Calsyn_Roades_1994}.

\subsubsection{Predictors about housing and services} \label{sec:housingservices} 

We observed algorithms in our corpus employed easily obtainable administrative data on \textbf{Housing} predictors (n=42, 72.4\%). These predictors again focused on ‘who’ clients are based on their (1) \textit{Service Usage History} detailing the type and length of shelter/homeless services an individual or family used previously (n=25, 43.1\%) and (2) \textit{Current Housing} and \textit{Housing History} detailing one's current/past living situation, eviction history, and any safety issues in their residence (n=22, 37.9\%). Fewer algorithms in our corpus included inputs that provided \textit{Service Provider Information} under \textbf{Services} (n=20, 34.5\%), which provides critical data on ‘what’ services (e.g., type and capacity of homeless service offered and the average wait time for services) are available to clients within homeless support systems. Moreover, even fewer algorithms in our dataset included \textit{Housing/service needs} (n=11, 19.0\%) inputs which signal how receptive a client is to receive homeless services (e.g., their willingness to wait for services and probability of success if partaking in services) and the service type/frequency a client needs. These predictors are hard to quantify or collect but can provide crucial insights into an individual's success in achieving stable housing. Because homeless support services often do not have enough resources to serve everyone \cite{Ecker2022, fowlersolving2019}, designing algorithms using quantifiable housing inputs that focus on the ‘who’ without equally considering (1) ‘what' their needs are and (2) if they are available, can render non-realistic algorithmic outputs. For example, even if an algorithm determines an individual is at high risk of experiencing chronic homelessness and should be placed in permanent housing, in reality, they may face a wait time of several years, which is not a feasible or useful output \cite{eubanks2018automating}.

\subsubsection{Predictors about person’s relationships, their needs, and risk factors}

Prior homelessness research and practitioners have advocated for strength-based approaches to support clients instead of case deficits \cite{Krabbenborg_Boersma_Wolf_2013, yvonne2012}. Yet, in our corpus, we observed algorithms included more deficit-related predictors. Strength-related predictors including, \textit{Relationship strengths} (n=7, 12.1\%), which inform whether individuals have family/friends they can rely on and \textit{Social network analysis} (n=8, 13.8\%), which contains social network data on the friends or social acquaintances of unhoused individuals were included in few algorithms. In contrast, we saw more inclusions of the \textbf{Person needs/Risks} dimension, which focuses on client risks (i.e., deficits). Codes under this dimension included \textit{Prior victimization/trauma} detailing prior experiences of abuse, neglect, or poverty (n=20, 34.5\%); \textit{Involvement with the criminal justice system} regarding prior arrests and incarcerations (n=16, 27.6\%) and \textit{Behavioral Characteristics} such as a youth’s attitudes and practices towards unprotected sex (n=2, 3.4\%). Moreover, 11 models (19.0\%) accounted for scores and responses in \textit{Risk Assessments} such as the Next Step Tool for unhoused youth \cite{chan_emp_19} and the Vulnerability Index-Service Prioritization Decision Assistance Tool (VI-SPDAT) \cite{kithulgoda_predictive_2022}. These risk assessments are widely-used triage tools used in federally mandated centralized homeless systems to assess one's vulnerability and prioritize services to high-needs clients \cite{homelesshub, nst}. However, in recent years, creators of the popular VI-SPDAT announced they were phasing out the tool due to its poor reliability, racial bias, and construct validity concerns \cite{orgcode, shinn22}. Our findings are thus important as they show many algorithms for homelessness are currently designed contrary to theory and practice. By focusing more on ‘what is wrong’ with a client (i.e., though a needs assessment), algorithms in our corpus are prone to falling into a deficit-cycle of creating negative expectations, experiences, and behaviors \cite{zimmer}. To design human-centered algorithms in homelessness, we should be considering strength-related predictors over deficits.

\begin{figure}[]
\centering 
\includegraphics[scale=0.55]{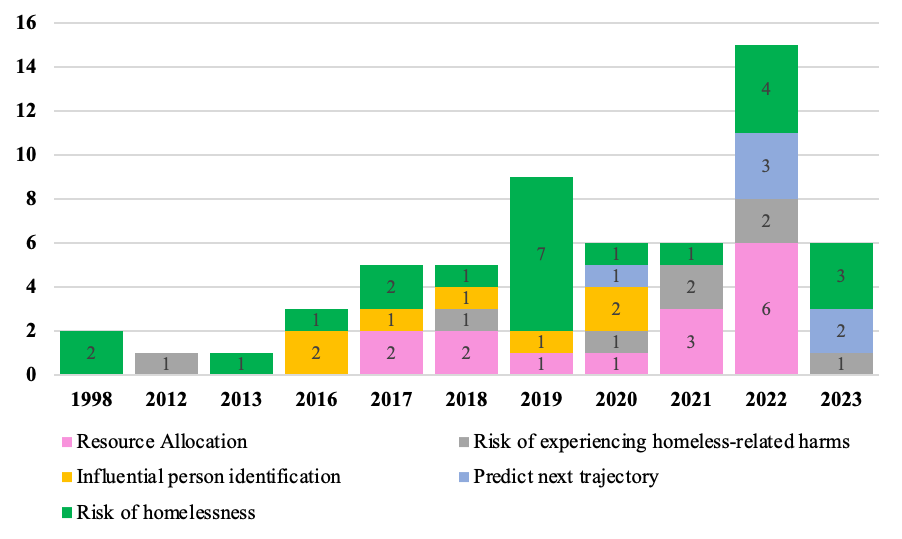}
\caption{Outcome variables for the algorithms}
\label{fig:outcome_time}
\end{figure}

\subsection{Outcome Variables in Algorithms for Homelessness} \label{sec:outcomevariable}

\subsubsection{Resource Allocation (RES)} \label{sec:resoutcome}

Resource allocation is a central problem of homelessness due to resource scarcity in the homeless system in many developed nations \cite{fowlersolving2019}. Fifteen algorithms (25.9\%) in our corpus were coded as resource allocation algorithms (RES). These algorithms were tasked to optimally allocate different resources (e.g., permanent supportive housing, rapid rehousing, counseling, psychiatric care) by matching unhoused individual's needs to available resources. RES algorithms in our datset also investigated how allocating resources based on different prioritization objectives (i.e., the most vulnerable or the most likely to take advantage of the services) can lead to different outcomes and impact our notions of fairness \cite{kube_just_2022}. Unsurprisingly, RES algorithms have increased in prominence in the last three years (see Figure \ref{fig:outcome_time}). This trend aligns with the adoption of federally-mandated centralized systems (i.e., coordinated systems) in Canada and the United States, which seek to efficiently match client needs to available resources through prioritization \cite{Ecker2022, azizi2018}. Recently, scholars have highlighted that such centralized systems may lead to inequitable access to housing due to (1) inherently subjective risk assessment (triage) tools that inaccurately assess someone's needs or (2) implementation challenges where data is missing or not up to date \cite{karusala19, kuo23}. In theory, RES algorithms could offer an opportunity to improve existing resource allocation systems to be fairer and more interpretable. However, as highlighted by our findings in section \ref{sec:ot}, we observed that RES algorithms in our corpus face external validity challenges because many were simulation studies.

\subsubsection{Predictive risk models: risk of experiencing homeless-related harms (HARM) and risk of homelessness (RISK)} \label{sec:conflate}

More than half the algorithms (n=31, 53.4\%) in our corpus were predictive risk models, predicting the risk of experiencing homelessness (RISK, n=23, 39.7\%) and homeless-related harms (HARM, n=8, 13.8\%). HARM algorithms predicted the risk of experiencing health issues resulting in emergency department visits, becoming a mental health inpatient, experiencing sustained homelessness, or substance abuse disorders. Many RISK models developed algorithms for specific subpopulation groups, including families that receive public assistance, veterans, youth, families, individuals who were previously homeless, and patients from substance abuse treatment programs. We found RISK and HARM models in our dataset appeared consistently over the years (see Figure \ref{fig:outcome_time}), peaking in 2019. Beyond the immediate role of predicting risk and harm, we observed these algorithms conflated risk assessment with resource allocation; where risk predictions were used to identify high needs clients to then prioritize these clients to receive support services. However, they differed fundamentally from the RES algorithms in our dataset by narrowly focusing on identifying ‘who’ is at high risk without considering `what' support services are available and how they would be assigned to individuals.

 \subsubsection{Influential person identification (ID) and predicting the next trajectory (TRJ)}

Target outcomes, identifying influential persons (ID, n=7, 12.1\%) and predicting the next trajectory (TRJ, n=6, 10.3\%) were the two least occurring algorithms in our dataset, emerging relatively recently since 2016 (see Figure \ref{fig:outcome_time}). These algorithms considered system-level components of the homeless system. All ID algorithms used network analysis approaches to simulate how to identify influential individuals among homeless youth to minimize violence and raise awareness about HIV. TRJ algorithms aimed to forecast or predict which homeless services will be assigned to a person based on their assessed level of need. With the implementation of centralized systems for resource allocation (i.e., coordinated systems), TRJ algorithms are increasingly relevant to understanding how individuals are assigned to different sequences of interventions \cite{Ecker2022}.

\subsection{Cross tabulation between outcome, predictors, and methods employed in algorithms for homelessness}

\subsubsection{Relationship between computational methods and outcome variables} \label{sec:mp} 

\begin{table}[]
\centering
\begin{tabular}{ll|ccccccccccc|}
\cline{3-13}
 &  & \multicolumn{11}{c|}{\textbf{Computational Method}} \\ \cline{3-13} 
 &  & \multicolumn{2}{c|}{\textbf{Inferential}} & \multicolumn{3}{c|}{\textbf{Machine}} & \multicolumn{1}{c|}{\textbf{Deep}} & \multicolumn{4}{c|}{\textbf{Optimization}} & \multicolumn{1}{l|}{} \\
 &  & \multicolumn{2}{c|}{\textbf{Statistic}} & \multicolumn{3}{c|}{\textbf{Learning}} & \multicolumn{1}{c|}{\textbf{Learning}} & \multicolumn{1}{l}{} & \multicolumn{1}{l}{} & \multicolumn{1}{l}{} & \multicolumn{1}{l|}{} & \multicolumn{1}{l|}{} \\ \cline{3-12}
 &  & \multicolumn{1}{c|}{\textbf{ST}} & \multicolumn{1}{c|}{\textbf{GLM}} & \multicolumn{1}{c|}{\textbf{SUP}} & \multicolumn{1}{c|}{\textbf{USUP}} & \multicolumn{1}{l|}{\textbf{NLP}} & \multicolumn{1}{c|}{\textbf{NN}} & \multicolumn{1}{l|}{\textbf{MIP}} & \multicolumn{1}{l|}{\textbf{NA}} & \multicolumn{1}{l|}{\textbf{SS}} & \multicolumn{1}{l|}{\textbf{HA}} & \multicolumn{1}{l|}{\textbf{Total}} \\ \hline
\multicolumn{1}{|l|}{\textbf{Outcome}} & \textbf{RES} & \multicolumn{1}{c|}{3} & \multicolumn{1}{c|}{5} & \multicolumn{1}{c|}{3} & \multicolumn{1}{c|}{1} & \multicolumn{1}{c|}{0} & \multicolumn{1}{c|}{0} & \multicolumn{1}{c|}{5} & \multicolumn{1}{c|}{0} & \multicolumn{1}{c|}{4} & \multicolumn{1}{c|}{2} & 23 \\
\multicolumn{1}{|l|}{\textbf{Variable}} & \textbf{HARM} & \multicolumn{1}{c|}{3} & \multicolumn{1}{c|}{6} & \multicolumn{1}{c|}{5} & \multicolumn{1}{c|}{0} & \multicolumn{1}{c|}{1} & \multicolumn{1}{c|}{1} & \multicolumn{1}{c|}{1} & \multicolumn{1}{c|}{0} & \multicolumn{1}{c|}{0} & \multicolumn{1}{c|}{0} & 17 \\
\multicolumn{1}{|l|}{\textbf{}} & \textbf{ID} & \multicolumn{1}{c|}{0} & \multicolumn{1}{c|}{0} & \multicolumn{1}{c|}{0} & \multicolumn{1}{c|}{0} & \multicolumn{1}{c|}{0} & \multicolumn{1}{c|}{0} & \multicolumn{1}{c|}{0} & \multicolumn{1}{c|}{7} & \multicolumn{1}{c|}{0} & \multicolumn{1}{c|}{0} & 7 \\
\multicolumn{1}{|l|}{\textbf{}} & \textbf{TRJ} & \multicolumn{1}{c|}{0} & \multicolumn{1}{c|}{2} & \multicolumn{1}{c|}{2} & \multicolumn{1}{c|}{0} & \multicolumn{1}{c|}{0} & \multicolumn{1}{c|}{2} & \multicolumn{1}{c|}{0} & \multicolumn{1}{c|}{1} & \multicolumn{1}{c|}{3} & \multicolumn{1}{c|}{0} & 10 \\
\multicolumn{1}{|l|}{\textbf{}} & \textbf{RISK} & \multicolumn{1}{c|}{1} & \multicolumn{1}{c|}{23} & \multicolumn{1}{c|}{10} & \multicolumn{1}{c|}{1} & \multicolumn{1}{c|}{0} & \multicolumn{1}{c|}{2} & \multicolumn{1}{c|}{2} & \multicolumn{1}{c|}{0} & \multicolumn{1}{c|}{0} & \multicolumn{1}{c|}{0} & 39 \\ \cline{2-13} 
\multicolumn{1}{|l|}{} & \textbf{Total} & \multicolumn{1}{c|}{7} & \multicolumn{1}{c|}{36} & \multicolumn{1}{c|}{20} & \multicolumn{1}{c|}{2} & \multicolumn{1}{c|}{1} & \multicolumn{1}{c|}{5} & \multicolumn{1}{c|}{8} & \multicolumn{1}{c|}{8} & \multicolumn{1}{c|}{7} & \multicolumn{1}{c|}{2} &  \\ \hline
\end{tabular}
\caption{Cross tabulation between the outcome variable and computational method}
\label{tab:cross-tab1}
\end{table}

From Table \ref{tab:cross-tab1}, we noticed RISK and HARM algorithms in our corpus predominantly employed GLMs and supervised ML approaches. Similarly, supervised machine learning models were often used to predict RISK (n=10) and HARM (n=5). These results point to the prevalence and focus on predicting risk in homelessness as a means to measure who is most at risk and should be prioritized \cite{Ecker2022}. We also saw that resource allocation algorithms (RES) used various optimization techniques, where mixed integer programming (MIP, n=5), as well as GLM methods (n=5), were most common. Our findings highlight the intrinsic limited utility of the RISK, HARM, and RES algorithms that employ GLMs and supervised ML– these models by design may follow ‘average trends,’ and ignore contextual circumstances impacting individuals.

\subsubsection{Relationship between computational methods and predictor variables} 

Cross-tabulating between computational methods and predictors (as seen in table \ref{tab:cross-tab2}) illuminated current limitations in homelessness algorithms. Models in our corpus using inferential statistics and ML techniques employed similar predictors: easily available predictors that fall under the \textbf{Demographics} and \textbf{Housing} dimensions to predict HARM and RISK. However, these models did not (if not rarely) used hard-to-quantify, \textit{Housing/service needs} predictors. As mentioned in \ref{sec:housingservices}, \textit{Housing/service needs} predictors inform the type and intensity of services a client needs and their receptiveness to receiving the service. Therefore, our findings showed that the inferential statistics and ML focused on ‘who’ a client is and failed to connect the ‘who’ to ‘what’ services they needed to achieve stable housing successfully. In contrast, we saw the reverse trend in algorithms using optimization approaches. These algorithms frequently used \textit{Service provider information} and \textit{Housing/service needs} to signal homeless resource supply-side information with less emphasis on who the client was.

\begin{table}[]
\centering
\resizebox{\columnwidth}{!}{%
\begin{tabular}{ll|cccc|ccccc}
 &  & \multicolumn{4}{c|}{\textbf{Computational Method}} & \multicolumn{5}{c}{\textbf{Outcome Variable}} \\ \cline{3-11} 
 &  & \multicolumn{1}{c|}{\textbf{Inferential}} & \multicolumn{1}{c|}{\textbf{Machine}} & \multicolumn{1}{c|}{\textbf{Deep}} & \textbf{Optimization} & \multicolumn{1}{c|}{\textbf{RES}} & \multicolumn{1}{c|}{\textbf{HARM}} & \multicolumn{1}{c|}{\textbf{ID}} & \multicolumn{1}{c|}{\textbf{TRJ}} & \textbf{RISK} \\
\multicolumn{2}{c|}{\textbf{Predictors}} & \multicolumn{1}{c|}{\textbf{Statistics (N=37)}} & \multicolumn{1}{c|}{\textbf{Learning (N=21)}} & \multicolumn{1}{c|}{\textbf{Learning (N=5)}} & \textbf{(N=24)} & \multicolumn{1}{c|}{\textbf{(N=15)}} & \multicolumn{1}{c|}{\textbf{(N=8)}} & \multicolumn{1}{c|}{\textbf{(N=7)}} & \multicolumn{1}{c|}{\textbf{(N=6)}} & \textbf{(N=23)} \\ \hline
\multicolumn{1}{l|}{\textbf{Demographics}} & Individual demographics (N=37) & \multicolumn{1}{c|}{32} & \multicolumn{1}{c|}{18} & \multicolumn{1}{c|}{4} & 6 & \multicolumn{1}{c|}{6} & \multicolumn{1}{c|}{8} & \multicolumn{1}{c|}{0} & \multicolumn{1}{c|}{1} & 23 \\ 
\multicolumn{1}{l|}{\textbf{}} & Household composition (N=11)& \multicolumn{1}{c|}{11} & \multicolumn{1}{c|}{5} & \multicolumn{1}{c|}{1} & 1 & \multicolumn{1}{c|}{0} & \multicolumn{1}{c|}{2} & \multicolumn{1}{c|}{0} & \multicolumn{1}{c|}{0} & 9 \\
\multicolumn{1}{l|}{\textbf{}} & Economic (N=24)& \multicolumn{1}{c|}{22} & \multicolumn{1}{c|}{11} & \multicolumn{1}{c|}{1} & 3 & \multicolumn{1}{c|}{2} & \multicolumn{1}{c|}{4} & \multicolumn{1}{c|}{0} & \multicolumn{1}{c|}{0} & 18 \\ \hline
\multicolumn{1}{l|}{\textbf{Housing}} & Current Housing (N=22)& \multicolumn{1}{c|}{19} & \multicolumn{1}{c|}{11} & \multicolumn{1}{c|}{3} & 3 & \multicolumn{1}{c|}{2} & \multicolumn{1}{c|}{1} & \multicolumn{1}{c|}{0} & \multicolumn{1}{c|}{3} & 16 \\
\multicolumn{1}{l|}{\textbf{}} & Housing History (N=22)& \multicolumn{1}{c|}{17} & \multicolumn{1}{c|}{10} & \multicolumn{1}{c|}{1} & 5 & \multicolumn{1}{c|}{3} & \multicolumn{1}{c|}{3} & \multicolumn{1}{c|}{0} & \multicolumn{1}{c|}{3} & 13 \\
\multicolumn{1}{l|}{\textbf{}} & Service usage history (N=25)& \multicolumn{1}{c|}{19} & \multicolumn{1}{c|}{9} & \multicolumn{1}{c|}{2} & 8 & \multicolumn{1}{c|}{6} & \multicolumn{1}{c|}{3} & \multicolumn{1}{c|}{0} & \multicolumn{1}{c|}{4} & 12 \\
\multicolumn{1}{l|}{\textbf{}} & Housing/service needs (N=11)& \multicolumn{1}{c|}{4} & \multicolumn{1}{c|}{3} & \multicolumn{1}{c|}{0} & 10 & \multicolumn{1}{c|}{10} & \multicolumn{1}{c|}{1} & \multicolumn{1}{c|}{0} & \multicolumn{1}{c|}{0} & 1 \\ \hline
\multicolumn{1}{l|}{\textbf{Services}} & Service provider information (N=20) & \multicolumn{1}{c|}{10} & \multicolumn{1}{c|}{6} & \multicolumn{1}{c|}{1} & 15 & \multicolumn{1}{c|}{12} & \multicolumn{1}{c|}{2} & \multicolumn{1}{c|}{0} & \multicolumn{1}{c|}{3} & 4 \\ \hline
\multicolumn{1}{l|}{\textbf{Health}} & Health (N=36)& \multicolumn{1}{c|}{34} & \multicolumn{1}{c|}{18} & \multicolumn{1}{c|}{3} & 5 & \multicolumn{1}{c|}{6} & \multicolumn{1}{c|}{8} & \multicolumn{1}{c|}{0} & \multicolumn{1}{c|}{0} & 23 \\  \hline
\multicolumn{1}{l|}{\textbf{Person Needs/}} & Prior victimization/trauma (N=20)& \multicolumn{1}{c|}{18} & \multicolumn{1}{c|}{9} & \multicolumn{1}{c|}{1} & 5 & \multicolumn{1}{c|}{4} & \multicolumn{1}{c|}{4} & \multicolumn{1}{c|}{0} & \multicolumn{1}{c|}{0} & 13 \\
\multicolumn{1}{l|}{\textbf{Risks}} & Involvement with criminal justice (N=16)& \multicolumn{1}{c|}{12} & \multicolumn{1}{c|}{5} & \multicolumn{1}{c|}{2} & 4 & \multicolumn{1}{c|}{3} & \multicolumn{1}{c|}{3} & \multicolumn{1}{c|}{2} & \multicolumn{1}{c|}{0} & 8 \\ 
\multicolumn{1}{l|}{\textbf{}} & Risk assessment (N=11)& \multicolumn{1}{c|}{10} & \multicolumn{1}{c|}{5} & \multicolumn{1}{c|}{1} & 2 & \multicolumn{1}{c|}{4} & \multicolumn{1}{c|}{1} & \multicolumn{1}{c|}{0} & \multicolumn{1}{c|}{0} & 6 \\
\multicolumn{1}{l|}{\textbf{}} & Behavioral characteristics (N=2) & \multicolumn{1}{c|}{2} & \multicolumn{1}{c|}{2} & \multicolumn{1}{c|}{1} & 1 & \multicolumn{1}{c|}{1} & \multicolumn{1}{c|}{2} & \multicolumn{1}{c|}{0} & \multicolumn{1}{c|}{0} & 0 \\ \hline
\multicolumn{1}{l|}{\textbf{Relationships}} & Relationship strengths (N=7)& \multicolumn{1}{c|}{7} & \multicolumn{1}{c|}{3} & \multicolumn{1}{c|}{0} & 1 & \multicolumn{1}{c|}{1} & \multicolumn{1}{c|}{2} & \multicolumn{1}{c|}{0} & \multicolumn{1}{c|}{0} & 5 \\
\multicolumn{1}{l|}{\textbf{}} & Social network analysis (N=8)& \multicolumn{1}{c|}{0} & \multicolumn{1}{c|}{1} & \multicolumn{1}{c|}{0} & 7 & \multicolumn{1}{c|}{0} & \multicolumn{1}{c|}{1} & \multicolumn{1}{c|}{7} & \multicolumn{1}{c|}{0} & 0
\end{tabular}%
}
\caption{Cross tabulation between predictors, outcome variables, and method }
\label{tab:cross-tab2}
\end{table}

\subsubsection{Relationship between target outcome and predictor variables} \label{sec:crosspanel}

Table \ref{tab:cross-tab2} shows the cross-tabulation between target outcomes and predictors. First, focusing on the most popular target outcome in our dataset, RISK models used predictors that inform \textit{Individual demographics} (n=23, 100\%), \textit{Health} (n=23, 100\%), \textit{Economic} (n=18, 78.3\%), and \textit{Current Housing} (n=16, 69.6\%). Surprisingly, few models employed predictors that signal risk factors: \textit{Prior victimization/trauma} (n=13, 56.5\%) and \textit{Involvement with criminal justice} (n=8, 34.8\%). And only one model utilized the \textit{Housing/services needs} (n=1, 4.35\%). Echoing earlier findings, we observed RISK algorithms unduly focused on using easily available individual-level demographics and housing information without sufficiently accounting for the specific needs of the unhoused person(s), raising external validity concerns. Finally, our examination of the least common target outcome, TRJ algorithms, showed that they reproduce the status quo. The models used a narrow set of predictors, \textit{Service usage history} (n=4, 66.7\%), \textit{Housing history} (n=3, 50\%), \textit{Current Housing} (n=3, 50\%), and \textit{Service provider information} (n=3, 50\%), which meant the models predicted TRJ based on prior data on typical service transition paths. 

\section{Discussion}

Our paper is situated within the SIGCHI research space that examines the implications, values, and perceptions of algorithms used in homelessness research \cite{showkat23, kuo23, karusala19}. At the same time, our work differs and adds to this area of study by using Baumer's HCAD lens to unveil how the technical underpinnings of these algorithms do not account for critical contextual information, are not grounded in reality, and reproduce the status quo. In this section, we problematize the design of these algorithms and propose opportunities for algorithmic design in homelessness research.

\subsection{Algorithms focus on prioritizing the \textit{who} without fully considering \textit{what} services are available or needed by clients} \label{sec:disc1}

The goal of the homeless system in many nations is to support unhoused individuals or those at risk of homelessness to end homelessness \cite{openingdoors, reachinghomes, australiahousing, nationalstrategies}. However, due to a growing unhoused population, demand for homeless support services often exceeds its supply \cite{Ecker2022, fowlersolving2019}. For example, many North American cities do not have enough shelter spaces or permanent housing available for the unhoused \cite{Ecker2022, fowlersolving2019}. To address this challenge, the US and Canada have introduced coordinated systems, which formulate the goal to end homelessness as a supply and demand problem through prioritization. Homeless systems aim to (1) identify those with the greatest need/vulnerability to determine ‘who’ should be prioritized to receive services and (2) determine what services they need. Finally, the systems aim to meet the prioritized individual’s needs (in other words, meet their \textit{demand}) by matching them with the available \textit{supply} of supportive resources.

Our results in Section \ref{sec:outcomevariable} show that \textbf{75.9\%} of algorithms in our dataset have the target outcomes codes, RISK, HARM, TRJ, and ID, which predict a person(s)' vulnerability or influence level to determine ‘who’ should be prioritized to receive sought-after homeless support services (\textbf{RQ3}). Our study problematizes this current dominant trend in homelessness research. Our results found the majority of these algorithms only consider a narrow and inflexible set of factors without considering the contextual circumstances and strengths of an individual. And most importantly, we learned these algorithms paid outsized attention to easily quantifiable, individual-level factors that only inform ‘who’ a client is while ignoring many nuanced indicators that specifically indicate \textit{what} the unhoused person(s) need to resolve their housing crisis (\textbf{RQ2}). For example, we saw this through simple numeric comparisons in Table \ref{tab:codebook_summary} where \textit{Individual demographics} were incorporated in \textbf{63.8\%} of algorithms while \textit{Service provider information} predictors were employed in \textbf{34.5\%} of algorithms. A possible reason for the prevalence of these `who' predictors may be because public sector algorithms are primarily built using large-scale administrative data \cite{mergel2016big, eubanks2018automating, saxena22}. This form of data, while easy to collect, offers "thin" macro-level insights into overarching trends at scale without providing "thick" contextual detail on the \textit{why} and \textit{how} \cite{aragon2022human, wang_thickthin}. Prior work by Showkat et al. \cite{showkat23} warn that homelessness algorithms which predominantly focus on identifying ‘who’ should be prioritized based on insensitive assumptions (e.g., taking gender as a time-invariant construct) can nullify a client’s contextual experiences and lead to biased outcomes. Our findings add support to but also extend these insights further as we highlight how “thin” insights into an individual’s circumstances not only renders limited assessments of a clients’ needs but can negatively and substantially impact the quality of services delivered to clients. Karusala et al. \cite{karusala19} highlight this limitation through ethnographic work. Through interviews with frontline social workers, the researchers found that because the homeless administrative database was designed to have one person to represent a family, caseworkers struggled to justify matching a family to a multi-bedroom unit because there was no information on their family members in the database. Additionally, our findings revealed a noticeable gap where hard-to-collect predictors that indicate a person's strength in overcoming homelessness (i.e., \textit{Social network analysis}, and \textit{Relationship Strengths}) were often excluded from models. Ignoring strength predictors goes against homelessness research which has found social ties with family members and relatives play a critical role in mitigating or precipitating the risk of homelessness and related harms \cite{pahwa2019, corinth2018, jurewicz2022}.

This current trend in identifying ‘who’ to prioritize based on easily obtainable, narrow insights becomes particularly problematic because many of these algorithms conflate needs/risk prediction with resource allocation; i.e., a client’s risk/harm predictions determine who gets prioritized to receive high-demand services such as permanent housing. Given the resource-constrained landscape within which the algorithms operate and because algorithmic predictions on prioritized individuals are inextricably tied to service delivery (see section \ref{sec:conflate}), we contend algorithms should be balancing identifying ‘who’ clients are with ‘what’ services are available to them to have any practical utility. Our findings are important because we show that by conflating algorithmic predictions with resource allocation, current algorithmic research on homelessness skip asking the important normative and value-laden questions: what counts as risk/harms; what types of risks should be prioritized, and for what services? Resonating with our findings, Kuo et al. previously interviewed stakeholders about their perceptions of an algorithm which prioritizes who should receive scarce housing resources. Study participants challenged the fundamental algorithmic notion that clients were required to show they are important enough to be provided housing support services \cite{kuo23}. Asking the above value-laden questions do not necessarily involve algorithmic interventions. Thus, instead of imagining and designing algorithmic interventions to mitigate the homelessness challenge, we propose researchers first question whether algorithms are needed in the first place \cite{baumer11}.

\subsection{Resource allocation algorithms are not grounded in real world conditions} \label{sec:disc2}

As noted in Section \ref{sec:disc1}, the US and Canada have federally mandated coordinated systems to intake, assess, and offer efficient systems of care to the unhoused \cite{Ecker2022}. Our results show that paralleling this movement, resource allocation algorithms (RES) in homelessness research have increased in frequency since 2017, becoming the most common outcome code in 2022 (see Section \ref{sec:resoutcome}). 

Perhaps due to their recent emergence, our findings show RES algorithms currently face external validity challenges. Notably, our findings showed the majority of RES algorithms use optimization approaches (\textbf{RQ1}) which can be perceived as hard-to-understand black-box systems that engender distrust of the homeless system because they generate inconsistent outcomes for similar person profiles \cite{chelmis_21, eubanks2018automating, karusala19}. Lack of explainability in algorithmic decisions and mistrust in these algorithms is problematic because this can leave clients feeling powerless against the system and discourage clients from sharing more information about their needs (e.g., prior trauma) \cite{karusala19, Showkat_Bellamy_To_2022}. Our findings revealed RES algorithms are also currently impractical as they are built on simulations based on theoretical population sizes (e.g., random range of values \cite{khayyatkhoshnevis_smart_2020}) and data drawn randomly from statistical distributions (e.g., resource capacity counts drawn from a uniform distribution \cite{chan_utilizing_2018}) (\textbf{RQ2}). While the simulations may work in theory, the algorithms can cause the \textit{ripple-effect trap} \cite{selbst2019fairness}, where the technology causes a slew of unintended consequences when deployed into existing homeless support systems and privilege easily quantifiable metrics over contextual observations made by caseworkers \cite{selbst2019fairness, chi23paper, raji22}. Moreover, close examination of the predictors in RES algorithms showed that the models included difficult-to-quantify and ill-defined predictors such as the risk level of youth binned into low, medium, and high \cite{chan_utilizing_2018} and an individual’s conflict value depicting their tendency for conflict \cite{saha2022}. No papers mentioned the practical challenges of quantifying these predictors even though prior works have found how we measure risk can be contested \cite{chi23paper, Ecker2022, jacobs20, saxena2022chilbw} and rapport with caseworkers and friends can influence behavioral predictor values \cite{karusala19,  van2014, Salem_Kwon_Ames_2018}. Instead, algorithms assumed these predictors were available and could correctly represent the construct of interest. By making such assumptions, algorithms risk falling into the \textit{solutionism trap} \cite{selbst2019fairness}, failing to recognize that technology may not be the best solution to the problem as the domain one is modeling for is so complex it is computationally intractable. It is also important to note that RES algorithms make the flawed deterministic assumption that an optimal solution for resource allocation exists (\textbf{RQ3}). For instance, a RES algorithm may conclude that an individual should enter an emergency shelter. However, this is often an impossible outcome for the estimated 8-25\% unhoused adults and youth population in the US and Canada that own pets \cite{french2021, mccosker2023}. With most shelters adopting a strict no-pets policy, prior works show unhoused pet owners refuse or are refused housing because of their pets \cite{kiddkidd94, lem2016, howe2018}. 

Guided by the limitations in RES algorithms, we offer concrete pathways for algorithmic research in homelessness. We argue future research should try to break away from current model-centric approaches that seek to optimize the parameters and performance of computational systems \cite{showkat23}. Instead, we recommend researchers consider shifting towards a data-centric AI paradigm (DCAI) \cite{dcai}, which magnifies the importance of data in computational models by critically inspecting data provenance and forging continuous engagements between AI and domain experts during the algorithm design and evaluation process. We also recommend adopting a human-centered participatory approach that involves key stakeholders to understand how structural (e.g., access to affordable housing and health services) and individual factors (e.g., relationships with family) dynamically impact hard-to-quantify facets of an individual, such as one’s resilience and willingness to overcome adversity. Only through these deep inquiries can we ascertain if decision-making technologies are computationally tractable in homelessness and further inquire which predictors and target outcomes should be incorporated in models \cite{hcai, aragon2022human, baumer2017toward, mothilal2024nonideal}.

\subsection{Algorithms should challenge the status quo instead of reproducing it} \label{sec:statusquo}
Our survey of literature showed that algorithms for homelessness reproduce the status quo of categorizing individuals and matching services to them based on their vulnerability. RISK and HARM models predominantly incorporated GLMs and ML techniques, which actively sustained the status quo by generating outcomes that follow average patterns identified from historical data to determine an individual's risk of homelessness or related harms (see Table \ref{tab:cross-tab1} and Section \ref{sec:rq1}) (\textbf{RQ1}). TRJ algorithms were built on a narrow set of historical data: what type of services other unhoused persons used and their homeless history to predict the next expected pathways for other individuals (see Section \ref{sec:crosspanel}) (\textbf{RQ2}). RES algorithms went further, using simulated data or pulling random instances from historical data (see Section \ref{sec:ot}) to allocate resources.

We should re-imagine algorithms for homelessness, and they should strive to challenge this status quo (\textbf{RQ3}). Extensive research on homelessness has found that commonly used triage tools and needs-based assessments focus on case deficits (i.e., ‘what is wrong with someone’) instead of strengths, leading to inequitable and biased outcomes, creating a deficit cycle and yielding incorrect scores \cite{karusala19, Ecker2022, cronley2022, osborne19, zimmer}. Even the creators of VI-SPDAT, the most commonly used triage tool in homelessness to prioritize services, have stopped supporting the use of their tool for these reasons \cite{shinn22, orgcode}. Yet our study findings show that research on algorithms for homelessness has predominantly focused on risk and harm prediction to prioritize service delivery. Given the marked heterogeneity among the unhoused population and their divergent needs \cite{fowlersolving2019}, we contend future research on homelessness algorithms should buck this trend and design holistic assessments that center unhoused client values and promote an individual's strengths instead of probabilistic models that produce singular risk scores. SIGCHI scholars can draw from HCI scholarship and adopt a \textit{speculative approach} \cite{baumer2017toward} to interrogate values, assumptions within homelessness algorithmic systems, and the ramification of design choices. For instance, Kuo et al. \cite{kuo23} took on this approach by introducing a comic boarding method to draw out frontline workers' and unhoused individuals' perspectives around a housing allocation algorithm implemented in the US. Saxena et al. \cite{saxena2021framework2} also offers inspiration for SIGCHI scholars on future pathways for designing more nuanced public sector decision-making algorithms. Introducing a theoretical framework for strength-based algorithmic decision-making that accounts for human discretion, algorithmic decision-making, and bureaucratic processes, the authors argue that public sector algorithms can reflexively focus on strengths (e.g., individual’s resiliency and strong ties with family that help them overcome adversity) through a prescriptive guiding framework to track and score a person’s circumstances across multiple domains.

\subsection{Heuristic Design Guidelines} 
From our results, we provide the following heuristic design guidelines for researchers along with examples on how to implement the guidelines: 
\begin {itemize} 
\item \textbf{Center impacted stakeholders in homelessness algorithm research to elucidate the boundaries of what computational models can do and ask whether we should be designing algorithms for this space \cite{muller2009participatory}:} Researchers can engage with stakeholders through ethnographic work, design workshops, and interview studies to deeply understand the tensions, perceptions, and limitations around computational interventions. Notably, questions around how to predict risk, who to prioritize, and who gets what resources are value-laden criteria that could be addressed prior to algorithmic interventions. 

\item \textbf{Empirically evaluate algorithms designed for homelessness and examine their external validity \cite{raji22, selbst2019fairness, coston2022}:} Researchers can adapt current resource allocation algorithms identified in our corpus using real-life datasets to evaluate their functionality, understand how they impact intersectional identities, and identify improvement areas.

\item \textbf{Adopt a human-centered theoretical approach to understand what data should or should not be used to inform decision-making \cite{saxena2021framework2} and articulate how identities and power relations are lost/uplifted through model building pipelines \cite{showkat23}:} Researchers can employ storytelling of individual stakeholder experiences and comic boarding methods \cite{raji22, muller2009participatory, kuo23} to empower stakeholders, encourage mutual and reciprocal learning, and fundamentally question taken-for-granted shortcomings in existing homelessness algorithms (such as the conflation of risk prediction with the prioritization of resource allocation).

\item \textbf{Use a speculative lens \cite{baumer2017toward} to engage with stakeholders to reimagine how to design flexible, strength-based systems that support collaborative decision-making and human discretion \cite{saxena2021framework2, raji22, chi23paper, robertson21}:} Researchers can draw on SIGCHI research from adjacent domains to imagine opportunities to ameliorate the algorithmic validity and functionality concerns we address in our study. For example, previously, Karusala et al. \cite{karusala19} found that caseworkers thought administrative datasets were missing important information and that algorithms should incorporate information from casenotes. In an adjacent domain, SIGCHI research has shown computational text analysis methodologies can unveil hard-to-capture street-level signals from child welfare casenotes but identified challenges in incorporating textual data into predictive model \cite{chi23paper, saxena22, field23}. Researchers can conduct similar analyses to explore how narratives can address current limitations in homelessness algorithms.

\item \textbf{Finally, remember the end goal:} The goal for designing and imagining futures for algorithms in homelessness research is to support unhoused clients get housing and stay housed. Researchers should not become distracted by model efficiency and output optimization and forget this mission.

\end{itemize}
\vspace{0.1cm}

\section{Limitations and future work}

Our systematic literature review was limited to our corpus of 57 articles. We may have missed algorithms detailed in other digital libraries or proprietary algorithms unavailable to the public for research. Moreover, our corpus predominantly focused on the US and Canadian context (n=55), all written in English. We plan to directly collaborate with homeless service providers and governmental agencies to deeply interrogate how algorithms shape the identities and experiences of impacted stakeholders. We will also navigate opportunities and limitations of implementing resource allocation algorithms through a co-design process with stakeholders to move towards designing human-centered algorithms.

\section{Conclusion}

We conducted a systematic literature review of 57 algorithms in homelessness research and deeply examined their computational methods, predictors and target outcomes using a human-centered lens \cite{baumer2017toward}. We examined trends and gaps in the current literature and provided future research opportunities. We call on the HCI community to actively engage with stakeholders in designing and evaluating algorithms proposed for homelessness. We recommend that future research in homeless algorithms explore how to build human-centered models that challenge the current practices that are strength-based, fair and context-aware.

\begin{acks}
This research was supported by the NSERC Discovery Early Career Researcher Grant RGPIN-2022-04570. Any opinions, findings, conclusions, and recommendations
expressed in this material are those of the authors. We are grateful for the anonymous reviewers whose suggestions
and comments helped improve the quality of this manuscript.
\end{acks}

\bibliographystyle{ACM-Reference-Format}
\bibliography{sample-base}

\appendix

\section{Appendix}

\begin{table}[H]
\centering
\begin{tabular}{c}
\hline
Combinations of Search Terms \\
\hline
"computation" AND (homelessness, OR homeless, OR unhoused) \\
"algorithm" AND (homelessness, OR homeless, OR unhoused) \\
"regression" AND (homelessness, OR homeless, OR unhoused) \\
"machine learning" AND (homelessness, OR homeless, OR unhoused)\\
"neural network" AND (homelessness, OR homeless, OR unhoused) \\
"ai" AND (homelessness, OR homeless, OR unhoused) \\
"resource allocation" AND (homelessness, OR homeless, OR unhoused) \\
"optimization" AND (homelessness, OR homeless, OR unhoused) \\
"housing insecurity" AND (homelessness, OR homeless, OR unhoused) \\
"data-driven" AND (homelessness, OR homeless, OR unhoused) \\ \hline
\end{tabular}
\caption{Combination of search terms used to search for articles}
\label{tab:searchterms}
\end{table}

\begin{figure}[H]
\centering 
\includegraphics[scale=0.5]{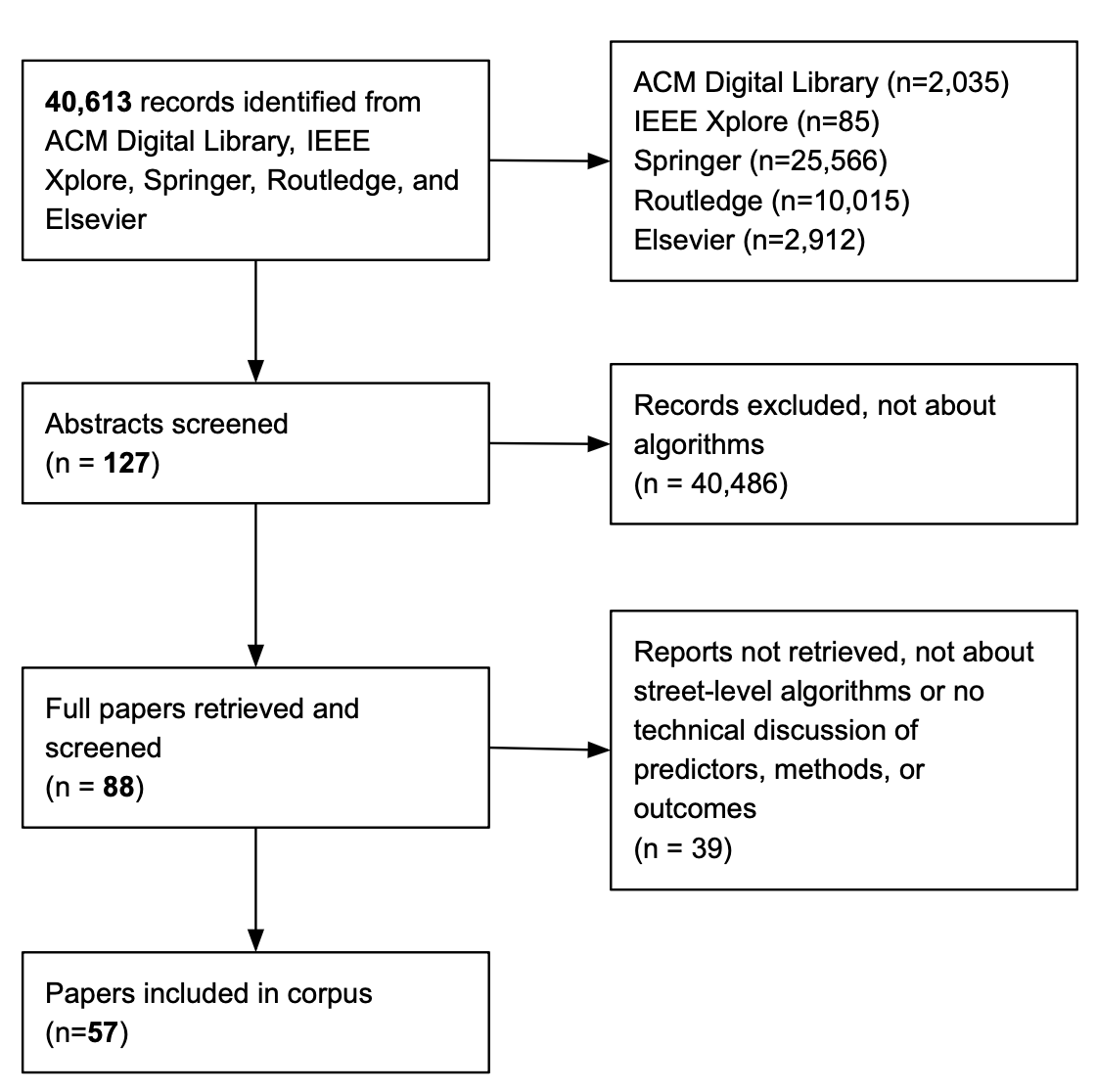}
\caption{Flow diagram of the literature review process}
\label{fig:prisma}
\end{figure}

\end{document}